\documentclass{elsart}
\usepackage{epsfig}
\usepackage{amssymb}

\begin{document}
\parindent=1em

\begin{frontmatter}

\title{Saturation of nuclear matter and radii of unstable nuclei} 

\author[a,b]{Kazuhiro Oyamatsu},
\author[b]{Kei Iida}

\address[a]{Department of Media Production and Theories, Aichi Shukutoku
University, Nagakute, Nagakute-cho, Aichi-gun, Aichi 480-1197, Japan}

\address[b]{The Institute of Physical and Chemical Research (RIKEN),
Hirosawa, Wako, Saitama 351-0198, Japan}

\begin{abstract}

     We examine relations among the parameters characterizing the 
phenomenological equation of state (EOS) of nearly symmetric, uniform nuclear 
matter near the saturation density by comparing macroscopic calculations of 
radii and masses of stable nuclei with the experimental data.  The EOS 
parameters of interest here are the symmetry energy $S_0$, the symmetry energy
density-derivative coefficient $L$ and the incompressibility $K_0$ at the 
normal nuclear density.  We find a constraint on the relation between $K_0$ 
and $L$ from the empirically allowed values of the slope of the saturation 
line (the line joining the saturation points of nuclear matter at finite 
neutron excess), together with a strong correlation between $S_0$ and $L$.  In
the light of the uncertainties in the values of $K_0$ and $L$, we 
macroscopically calculate radii of unstable nuclei as expected to be produced 
in future facilities.  We find that the matter radii depend strongly on $L$ 
while being almost independent of $K_0$, a feature that will help to determine
the $L$ value via systematic measurements of nuclear size.

\end{abstract}
\begin{keyword}
  Dense matter \sep Saturation \sep Unstable nuclei
\PACS 21.65.+f \sep  21.10.Gv
\end{keyword}
\end{frontmatter}


\section{Introduction}
\label{sec:intro}

    Saturation of density and binding energy, one of the fundamental 
properties of atomic nuclei, underlies a liquid-drop approach described by
a Weizs{\" a}cker-Bethe mass formula \cite{BW}
\begin{equation}
    -E_B=E_{\rm vol}+E_{\rm sym}+E_{\rm surf}+E_{\rm Coul},
\label{wb}
\end{equation}
where $E_B$ is the nuclear binding energy, $E_{\rm vol}$ is the volume energy,
$E_{\rm sym}$ is the symmetry energy, $E_{\rm surf}$ is the surface energy, 
and $E_{\rm Coul}$ is the Coulomb energy.  The sum $E_{\rm vol}+E_{\rm sym}$
corresponds to the saturation energy of uniform nuclear matter.  Since matter 
in nuclei is a strongly interacting system, it remains a challenging 
theoretical problem to understand the nuclear matter equation of state (EOS) 
through microscopic calculations that utilize a model of the nuclear force 
duly incorporating low-energy two-nucleon scattering data and properties of 
light nuclei \cite{HP}.  Furthermore, it is not straightforward to empirically
clarify the EOS, although constraints on the EOS from nuclear masses and radii
(e.g., Refs.\ \cite{OTSST,B,CWS}), observables in heavy-ion collision 
experiments performed at intermediate and relativistic energies (e.g., Refs.\ 
\cite{Li,Sturm,FFZZ,Daniel}), the isoscalar giant monopole resonance in nuclei
(e.g., Ref.\ \cite{YCL}) and even X-ray observations of isolated neutron stars
\cite{Pons} are available.  In this work we will consider such constraints 
from nuclear masses and radii.

    The EOS of bulk nuclear matter is a function of nucleon density $n$ and 
proton fraction $x$, which are related to the neutron and proton number 
densities $n_n$ and $n_p$ as $n_n=n(1-x)$ and $n_p=nx$.  We may generally 
express the energy per nucleon near the saturation point of symmetric nuclear
matter as \cite{L}
\begin{equation}
    w=w_0+\frac{K_0}{18n_0^2}(n-n_0)^2+ \left[S_0+\frac{L}{3n_0}(n-n_0)
      \right]\alpha^2.
\label{eos0}
\end{equation}
Here $w_0$, $n_0$ and $K_0$ are the saturation energy, the saturation density 
and the incompressibility of symmetric nuclear matter, $S_0$ is the symmetry 
energy $S(n)$ at $n=n_0$, $L=3n_0(dS/dn)_{n=n_0}$ is the density symmetry
coefficient, and $\alpha=1-2x$ is the neutron excess.  As the neutron excess 
increases from zero, the saturation point moves in the density versus energy 
plane.  This movement is determined mainly by the parameters $L$ and $S_0$ 
associated with the density-dependent symmetry energy $S(n)$ \cite{OTSST}.  Up
to second order in $\alpha$, the saturation energy $w_s$ and density $n_s$ are
given by 
\begin{equation}
  w_s=w_0+S_0 \alpha^2
\label{ws}
\end{equation}
and
\begin{equation}
  n_s=n_0-\frac{3 n_0 L}{K_0}\alpha^2.
\label{ns}
\end{equation}
The slope, $y$, of the saturation line near $\alpha=0$ $(x=1/2)$ is thus 
expressed as
\begin{equation}
 y=-\frac{K_0 S_0}{3 n_0 L}.
\label{slope}
\end{equation}

    Derivation of $L$ and $S_0$ from nuclear observables is generally obscured
by the interfacial and electrostatic properties.  Among the observables, the 
masses and root-mean-square radii of nuclei, which are controlled mainly by 
the bulk properties, are expected to be good tracers of $L$ and $S_0$.  This 
expectation was stressed by an earlier investigation \cite{OTSST} based on two
macroscopic nuclear models.  Such macroscopic models are reliable in a range 
of neutron excess, $\alpha\lesssim0.3$, and mass number, $A\gtrsim50$.  In 
this range the neutron separation energy that can be evaluated from a 
Weizs{\" a}cker-Bethe mass formula is greater than 2 MeV, allowing us to 
preclude the possibility of neutron halo formation expected at large neutron 
excess (or small separation energy).  In constraining the EOS from masses and 
radii of nuclei of neutron excess $\alpha\lesssim0.3$ and mass number 
$A\gtrsim50$ within the framework of the macroscopic models, systematic study 
allowing for uncertainties in the EOS parameters is indispensable.

    In this paper we thus explore a systematic way of extracting $L$ and $S_0$
from empirical masses and radii of nuclei, together with the parameters, 
$n_0$, $w_0$ and $K_0$, characterizing the saturation of symmetric nuclear 
matter.  We first set an expression for the energy of uniform nuclear matter, 
which reduces to the phenomenological form (\ref{eos0}) in the limits of $n\to
n_0$ and $\alpha\to0$ $(x\to 1/2)$.  Using this energy expression within a 
simplified version of the extended Thomas-Fermi approximation, which permits 
us to determine the macroscopic features of the nuclear ground state, we 
calculate charges, charge radii and masses of $\beta$-stable nuclei.  Comparing
these calculations with empirical values allows us to derive the optimal 
parameter set for various values of the slope $y$ and the incompressibility 
$K_0$.  We thus find a strong correlation between $L$ and $S_0$.  The next 
step is to calculate root-mean-square charge and matter radii of more 
neutron-rich nuclei that are expected to be produced in future radioactive ion
beam facilities.  The results suggest that the density symmetry coefficient 
$L$ may be constrained from possible systematic data on the matter radii in a 
way nearly independent of the poorly known $K_0$, while the slope $y$ being 
deducible as a function of $K_0$.  We finally discuss an empirically allowed 
region in the space of the parameters characterizing the EOS (\ref{eos0}).

     In Section \ref{sec:macro} we construct a macroscopic model of nuclei. 
Optimization associated with fitting to empirical data for nuclei on the 
smoothed $\beta$-stability line is illustrated in Section \ref{sec:opt}.  In 
Section \ref{sec:radii} we calculate matter and charge radii of unstable 
nuclei.  Our conclusions are presented in Section \ref{sec:conc}.  Numerical 
tables are given in the Appendix.

\section{Macroscopic description of nuclei}
\label{sec:macro}

    In constructing a macroscopic nuclear model, we begin with the expression 
for the bulk energy per nucleon \cite{O}, 
\begin{equation}
  w=\frac{3 \hbar^2 (3\pi^2)^{2/3}}{10m_n n}(n_n^{5/3}+n_p^{5/3})
      +(1-\alpha^2)v_s(n)/n+\alpha^2 v_n(n)/n,
\label{eos1}
\end{equation}
where 
\begin{equation}
  v_s=a_1 n^2 +\frac{a_2 n^3}{1+a_3 n}
\label{vs}
\end{equation}
and
\begin{equation}
  v_n=b_1 n^2 +\frac{b_2 n^3}{1+b_3 n}
\label{vn}
\end{equation}
are the potential energy densities for symmetric nuclear matter and pure 
neutron matter, and $m_n$ is the neutron mass. (Replacement of the proton mass
$m_p$ by $m_n$ in the proton kinetic energy makes only a negligible 
difference.)  For the later purpose of roughly describing the nucleon 
distribution in a nucleus, we incorporate into the potential energy densities 
(\ref{vs}) and (\ref{vn}) a low density behaviour $\propto n^2$ as expected 
from a contact two-nucleon interaction.  Note, however, that we will focus on 
the EOS of nearly symmetric nuclear matter near the saturation density.  We 
will thus determine the parameters included in Eqs.\ (\ref{vs}) and (\ref{vn})
in such a way that they reproduce data on radii and masses of {\em stable} 
nuclei.

     In the limit of $n\to n_0$ and $\alpha\to0$ $(x\to1/2)$, expression 
(\ref{eos1}) reduces to the usual form (\ref{eos0}) according to
\begin{equation}
 S_0= \frac16 \left(\frac{3\pi^2}{2}\right)^{2/3}\frac{\hbar^2}{m_n}n_0^{2/3}
  +(b_1-a_1)n_0+\left(\frac{b_2}{1+b_3 n_0}-\frac{a_2}{1+a_3 n_0}\right)n_0^2,
\label{s0}
\end{equation}
\begin{eqnarray}
 \frac13 n_0 L
 &=&\frac19\left(\frac{3\pi^2}{2}\right)^{2/3}\frac{\hbar^2}{m_n}n_0^{5/3}
     +(b_1-a_1)n_0^2
 +2\left(\frac{b_2}{1+b_3 n_0}-\frac{a_2}{1+a_3 n_0}\right)n_0^3
\nonumber \\ & &  
     -\left[\frac{b_2 b_3}{(1+b_3 n_0)^2}
           -\frac{a_2 a_3}{(1+a_3 n_0)^2}\right]n_0^4,
\label{p0}
\end{eqnarray}
\begin{equation}
 w_0=\frac{3}{10}\left(\frac{3\pi^2}{2}\right)^{2/3}\frac{\hbar^2}{m_n}
     n_0^{2/3} + a_1 n_0+\frac{a_2 n_0^2}{1+a_3 n_0},
\label{w0}
\end{equation}
\begin{equation}
 K_0=-\frac35\left(\frac{3\pi^2}{2}\right)^{2/3}\frac{\hbar^2}{m_n}n_0^{2/3}
     +\frac{18 a_2 n_0^2}{(1+a_3 n_0)^3},
\label{k0}
\end{equation}
\begin{equation}
 0=\frac15\left(\frac{3\pi^2}{2}\right)^{2/3}\frac{\hbar^2}{m_n}n_0^{-1/3}
     +a_1+\frac{2a_2 n_0}{1+a_3 n_0}-\frac{a_2 a_3 n_0^2}{(1+a_3 n_0)^2}.
\label{sat}
\end{equation}
Equation (\ref{sat}) comes from the fact that the density derivative of the 
energy per nucleon at $n=n_0$ and $\alpha=0$ $(x=1/2)$ vanishes due to the 
saturation.  Note that the five relations (\ref{s0})--(\ref{sat}) are not 
sufficient to determine the six parameters $a_1, \ldots, b_3$, and that the 
parameter $b_3$, which controls the EOS of matter at large neutron excess and 
high density, has little effect on the saturation properties of nearly 
symmetric nuclear matter.  We will thus set $b_3$ as a typical value 1.58632 
fm$^3$, which was obtained by one of the authors \cite{O}, and determine the 
other parameters from the empirical radii and masses of stable nuclei.

    We now proceed to describe a spherical nucleus of proton number $Z$ and 
mass number $A$ within the framework of a simplified version of the extended 
Thomas-Fermi theory \cite{O}.  We first write the total energy of a nucleus 
as a function of the density distributions $n_n({\bf r})$ and $n_p({\bf r})$ 
according to
\begin{equation}
 E=E_b+E_g+E_C+Nm_n+Zm_p,
\label{e}
\end{equation}
where 
\begin{equation}
  E_b=\int d^3 r n({\bf r})w\left(n_n({\bf r}),n_p({\bf r})\right)
\label{eb}
\end{equation}
is the bulk energy,
\begin{equation}
  E_g=F_0 \int d^3 r |\nabla n({\bf r})|^2
\label{eg}
\end{equation}
is the gradient energy with an adjustable constant $F_0$,
\begin{equation}
  E_C=\frac{e^2}{2}\int d^3 r \int  d^3 r' 
      \frac{n_p({\bf r})n_p({\bf r'})}{|{\bf r}-{\bf r'}|}
\label{ec}
\end{equation}
is the Coulomb energy, and $N=A-Z$ is the neutron number.  Here we ignore 
shell and pairing effects.  We also neglect the contribution to $E_g$ from the
gradient of the proton fraction $x$ [see Eq.\ (\ref{eg})]; this contribution 
makes only a little difference even in the description of extremely 
neutron-rich nuclei, as clarified in the context of neutron star matter 
\cite{O}.

   The energy (\ref{e}), once optimized, can be mapped onto a 
Weizs{\" a}cker-Bethe mass formula of the form (\ref{wb}) via
\begin{equation}
  E_{\rm surf}=E_g+(E_b-E_{\rm vol})+(E_C-E_{\rm Coul}),
\label{esurf}
\end{equation}
where $E_b-E_{\rm vol}$ denotes the inhomogeneity contribution to the bulk 
energy $E_b$, and $E_C-E_{\rm Coul}$ ($E_{\rm Coul}=3 Z^2 e^2/5R$ with the 
liquid-drop radius $R$) denotes that to the Coulomb energy $E_C$.  These 
inhomogeneity contributions arise from the fact that matter in a nucleus is 
compressible.  We remark that in equilibrium with respect to nuclear size, 
$E_g=E_C$ holds \cite{O}.  Combining this relation with the well-known 
equilibrium condition for the liquid-drop size, $E_{\rm surf}=2E_{\rm Coul}$, 
and with $E_{\rm Coul}\simeq E_C$, we find a simple relation, $E_g\simeq 
E_{\rm surf}/2$, for an equilibrium nuclide.

    For the present purpose of examining the macroscopic properties of nuclei 
such as masses and radii, it is sufficient to characterize the neutron and 
proton distributions for each nucleus by the central densities, radii and 
surface diffuseness different between neutrons and protons, as in Ref.\ 
\cite{O}.  We thus assume the nucleon distributions $n_i(r)$ $(i=n,p)$, where 
$r$ is the distance from the center of the nucleus, as
\begin{equation}
  n_i(r)=\left\{ \begin{array}{lll}
  n_i^{\rm in}\left[1-\left(\displaystyle{\frac{r}{R_i}}\right)^{t_i}\right]^3,
         & \mbox{$r<R_i,$} \\
             \\
         0\ ,
         & \mbox{$r\geq R_i.$}
 \end{array} \right.
\label{ni}
\end{equation}
Here $R_i$ roughly represents the nucleon radius, $t_i$ the relative surface 
diffuseness, and $n_i^{\rm in}$ the central number density.  The proton 
distribution of the form (\ref{ni}) can fairly well reproduce the experimental
data for stable nuclei such as $^{90}$Zr and $^{208}$Pb, as we shall see in 
the next section.

\begin{figure}
\begin{center}
\epsfig{file=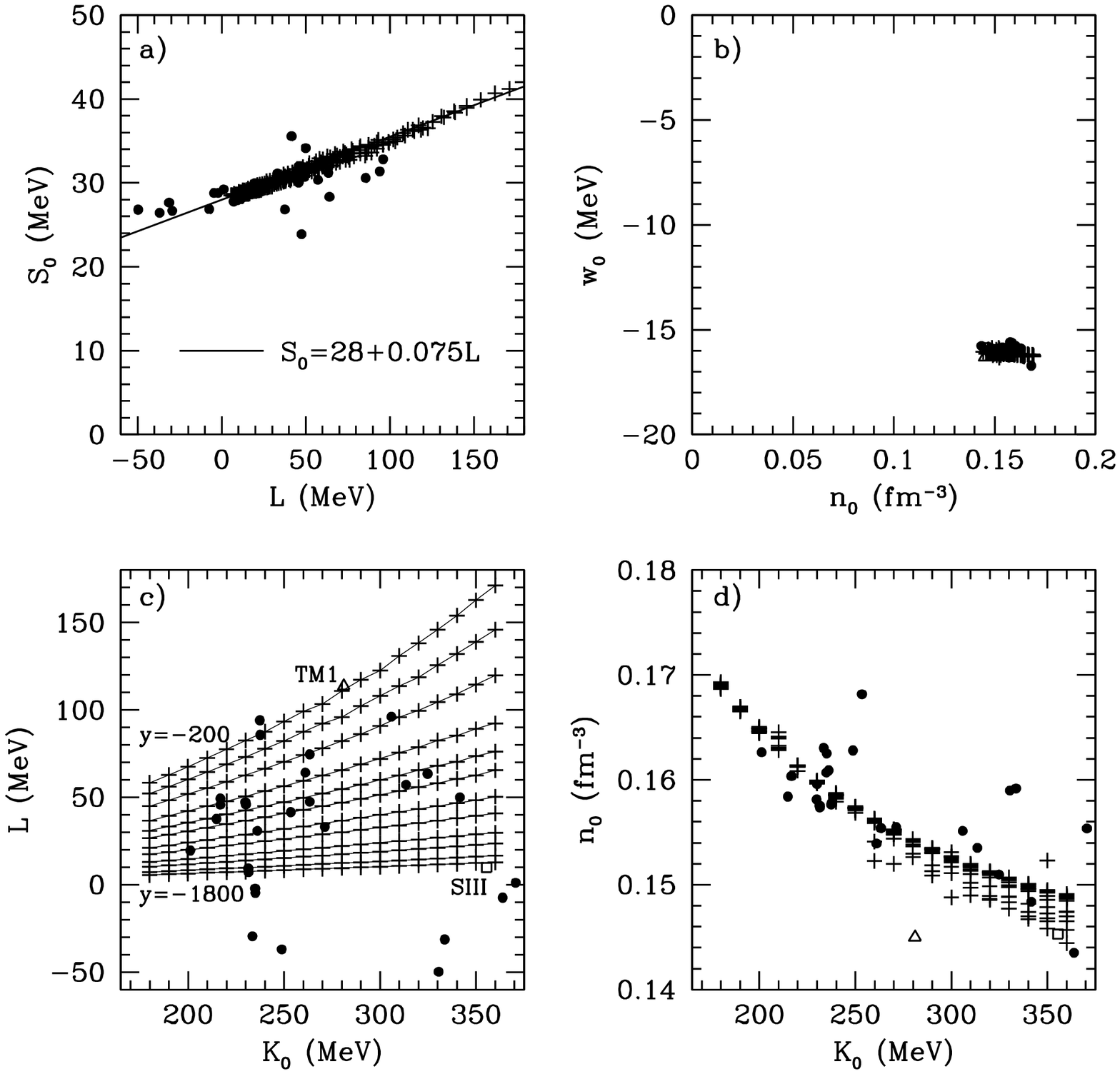,height=13cm}
\end{center}

\begin{flushleft}
{\small
Fig.\ 1.  Various optimal relations among the parameters $S_0$, $n_0$, $w_0$, 
$L$ and $K_0$ characterizing the EOS of nearly symmetric nuclear matter.  In
addition to the present results (crosses), the Skyrme-Hartree-Fock predictions
[dots except for SIII (square)] and the TM1 prediction (triangle) are plotted.
In (c), the thin lines are lines of constant $y$.}
\end{flushleft}

\end{figure}

\section{Optimization}
\label{sec:opt}

    For fixed mass number $A$, we then optimize the energy (\ref{e}) with 
respect to the parameters $R_i$, $t_i$ and $n_i^{\rm in}$.  (In a general 
Thomas-Fermi approach, such optimization is carried out without assuming a 
particular form of the distributions.)  The resultant optimal values of charge
number, nuclear mass and root-mean-square charge radius
\begin{equation}
    R_c=\left[Z^{-1}\int d^3 r r^2 \rho_c({\bf r})\right]^{1/2},
\label{Rc}
\end{equation}
where 
\begin{equation}
    \rho_c({\bf r})=(\pi^{1/2}a_p)^{-3}\int d^3 r' 
                  \exp\left({-|{\bf r}-{\bf r'}|^2/a_p^2}\right)n_p({\bf r'})
\label{rhoc}
\end{equation}
with $a_p=0.65$ fm is the charge distribution folded with the proton form 
factor \cite{ES}, are functions of $a_1$--$b_3$ and $F_0$.  These optimal 
values are in turn compared with the empirical values for nuclei on the 
smoothed $\beta$-stability line ranging $25\leq A \leq 245$ (see Table A.1 in 
Ref.\ \cite{O}, which is based on Refs.\ \cite{Y,VJV}).  For fixed slope $y$ 
and incompressibility $K_0$, such a comparison can be made by a usual least 
squares fitting, which gives rise to an optimal set of the parameters 
$a_1$--$b_3$ and $F_0$.  Here, we set $y$ and $K_0$ as $-1800$ MeV 
fm$^3 \leq y \leq -200$ MeV fm$^3$ and 180 MeV $\leq K_0 \leq 360$ MeV; the 
numerical results for $n_0$, $w_0$, $S_0$, $L$ and $F_0$ are tabulated in the 
Appendix.  When the slope is very gentle ($0> y\gtrsim -200$ MeV fm$^3$), the 
optimal set is unavailable for large $K_0$; when the slope is very steep, the 
optimal parameters converge on the values close to those obtained for 
$y=-1800$ MeV fm$^3$.  Nuclear masses that can be calculated from the optimal 
parameter sets agree well with the experimental data of 1962 nuclides 
$(A\geq2)$ \cite{AW}; for all the combinations of $y$ and $K_0$, the 
root-mean-square deviations of the masses are $\sim3$--5 MeV, which are 
roughly as large as those obtained from a Weizs{\" a}cker-Bethe type formula.
We likewise evaluated the root-mean-square charge radii of various stable 
nuclei; the root-mean-square deviations from the experimental data of 92 
nuclides $(A\geq50)$ \cite{VJV} are about 0.06 fm.

\begin{figure}
\begin{center}
\epsfig{file=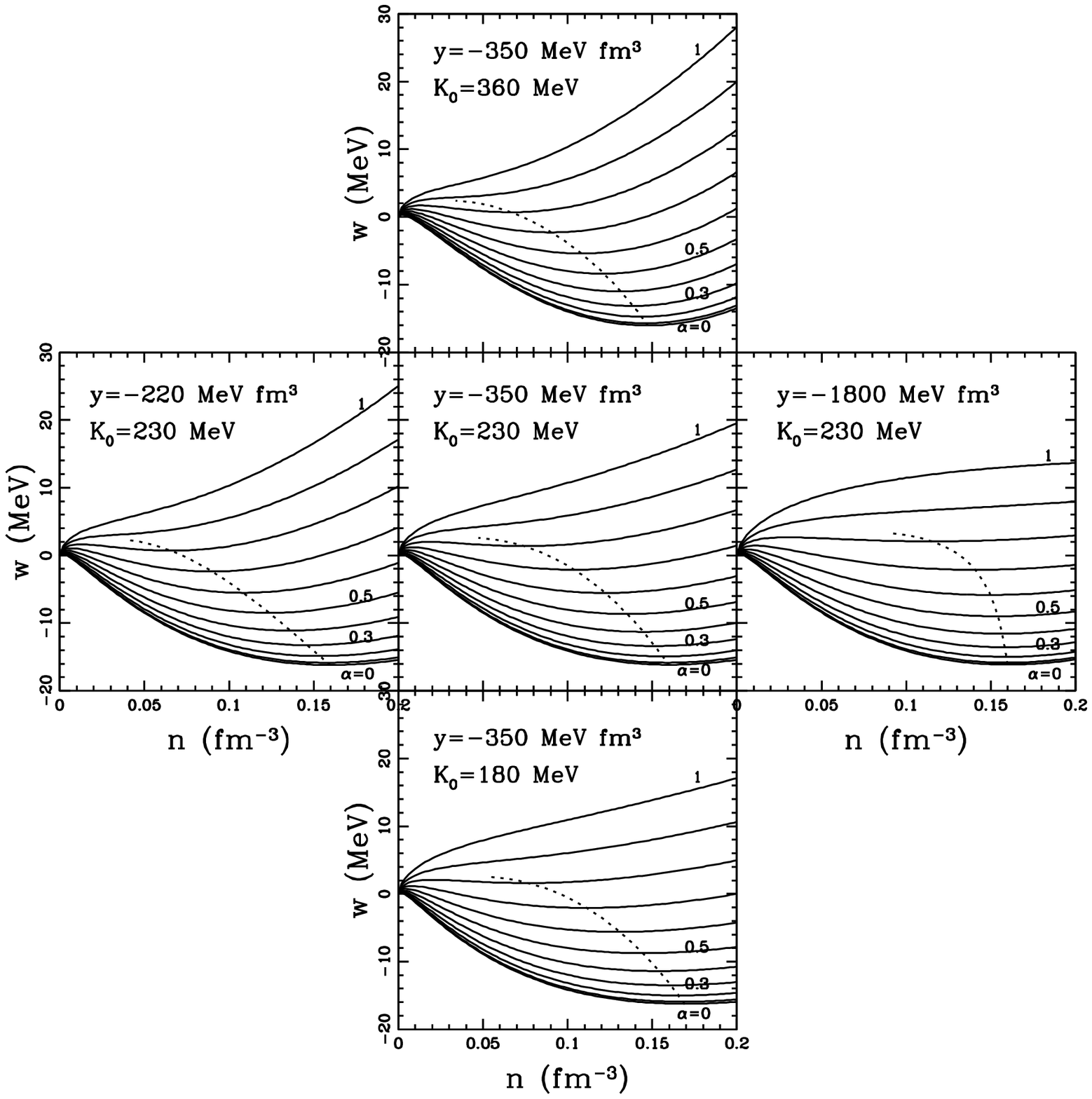,height=13cm}
\end{center}

\begin{flushleft}
{\small
Fig.\ 2.  The energy per nucleon as a function of nucleon density.  The dotted
lines denote the saturation lines.} 
\end{flushleft}

\end{figure}

    The optimal results for $S_0$, $L$, $n_0$ and $w_0$ are plotted in Fig.\ 1.
From Fig.\ 1a, we obtain a relation nearly independent of $K_0$:
\begin{equation}
    S_0 \approx B+C L,
\label{s0p0}
\end{equation}
where $B\approx 28$ MeV and $C\approx 0.075$.  We find from Figs.\ 1b and 1d 
that the saturation energy and density of symmetric nuclear matter always take
on a value of $-16.0\pm0.5$ MeV and $0.155\pm0.015$ fm$^{-3}$.  Several data in
Fig.\ 1d having $n_0$ smaller than the standard values correspond to the case 
of $y=-200$ MeV fm$^3$, where the fitting is no longer effective.  In Fig.\ 1c,
we see a band on which the optimal values of $L$ and $K_0$ are scattered; in 
this band $L$ increases with increasing $y$ for fixed $K_0$.  For comparison 
we also plot the predictions from various Skyrme-Hartree-Fock schemes 
(references in Ref.\ \cite{B}; Refs.\ \cite{GTP,SGHPT,CBHMS}) and a 
relativistic mean field model (TM1 in Ref.\ \cite{ST}); the values for $n_0$, 
$w_0$, $K_0$, $S_0$, $L$ and $y$ are tabulated in Table 1.  These predictions 
are distributed over the band, among which only two (TM1 and SIII) were 
considered in the previous analysis \cite{OTSST}.  We remark in passing that 
the optimal values of $F_0$ are confined to $66\pm6$ MeV fm$^5$, consistent 
with the result of Ref.\ \cite{O}.

     In Fig.\ 2 we display the EOS (\ref{eos1}) for various sets of $y$ and 
$K_0$.  Whereas $y$ affects the slope of the saturation line, $K_0$ controls
$n_0$ as well as the curvature of the line of constant $\alpha$.  The neutron 
and proton distributions in $^{208}$Pb and $^{90}$Zr modelled via Eq.\ 
(\ref{ni}) are plotted in Fig.\ 3.  The question we consider in the next 
section is how such differences in the saturation properties as shown in Fig.\
2 affect matter and charge radii of unstable nuclei that can be evaluated from
the distributions of the form illustrated in Fig.\ 3.

\begin{figure}
\begin{center}
\epsfig{file=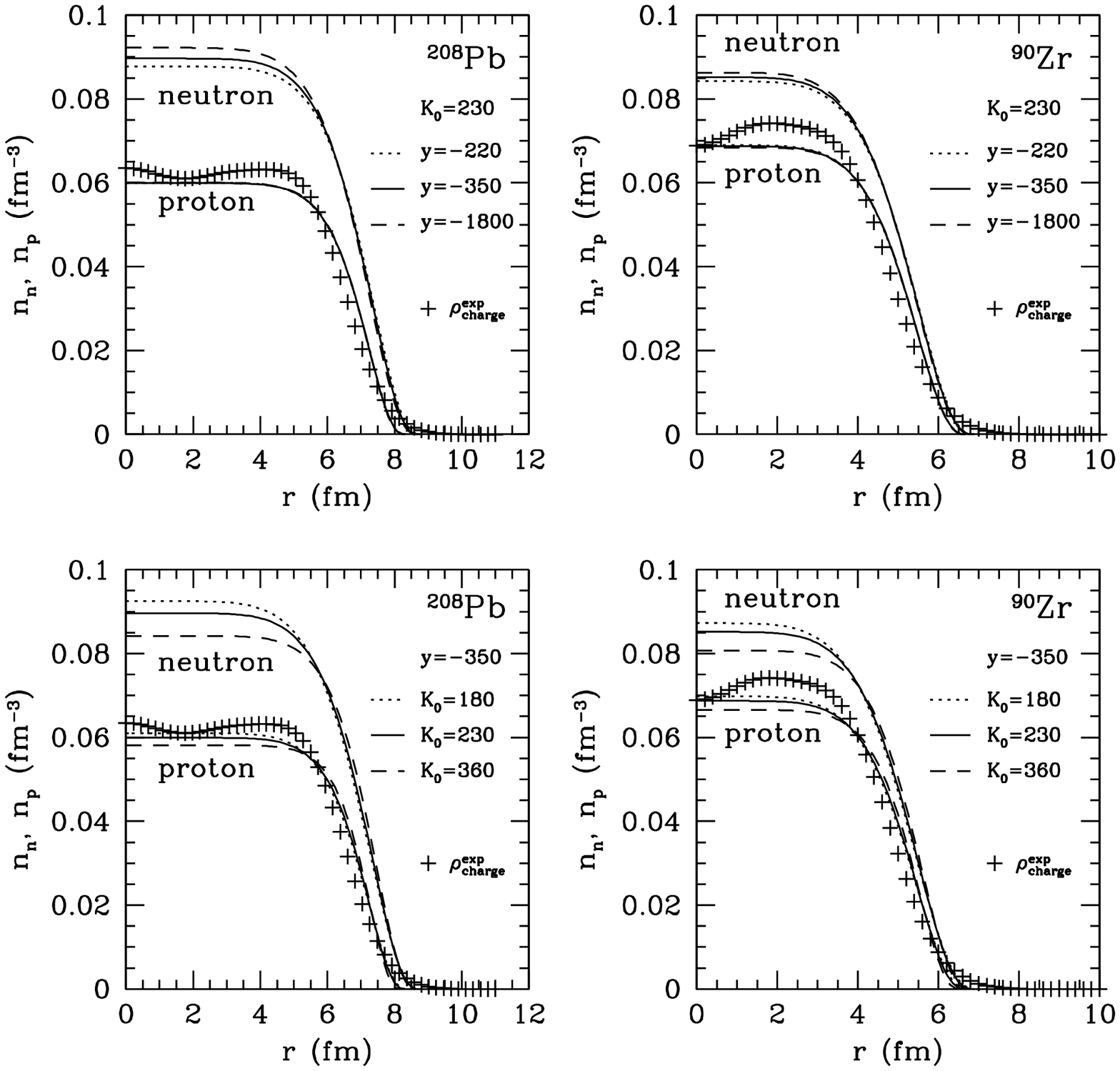,height=13cm}
\end{center}

\begin{flushleft}
{\small
Fig.\ 3.  The neutron and proton density distributions in $^{208}$Pb and 
$^{90}$Zr.  Experimental data on the charge distributions (crosses) are taken
from Ref.\ \cite{VJV}.}
\end{flushleft}

\end{figure}

\section{Matter and charge radii}
\label{sec:radii}

    For neutron-rich nuclides we now obtain the root-mean-square charge radii
$R_c$ and matter radii $R_m$, defined as
\begin{equation}
    R_m=\left[A^{-1}\int d^3 r r^2 \rho_m({\bf r})\right]^{1/2},
\label{Rm}
\end{equation}
where
\begin{equation}
    \rho_m({\bf r})=(\pi^{1/2}a_p)^{-3}\int d^3 r' 
                     \exp\left({-|{\bf r}-{\bf r'}|^2/a_p^2}\right)n({\bf r'})
\label{rhom}
\end{equation}
is the matter distribution folded with the proton charge form factor equally
for neutrons and protons.  We evaluated the radii $R_c$ and $R_m$ 
of Ni and Sn isotopes for combinations of $y=-220,-350,-1800$ MeV fm$^3$ and 
$K_0=180,230,360$ MeV.  The results are shown at neutron excess of up to 
$\alpha\sim0.4$ in Fig.\ 4.  We find from the upper panels of Fig.\ 4 (see 
also the upper panels of Fig.\ 5 for further clarity) that at $\alpha\sim0.3$ 
$(x\sim0.35)$ a difference of order 0.05--0.1 fm occurs in the matter radii 
due to variation in the slope $y$.  This is because as the slope becomes 
gentler, the saturation density difference $n_0-n_s$ becomes larger [see Eq.\ 
(\ref{ns})].  We also find that the charge radii depend only weakly on $y$, a 
feature consistent with the prediction made in Ref.\ \cite{OTSST}.  We can 
thus expect that forthcoming empirical data on matter radii of unstable 
neutron-rich nuclei with accuracy down to order $\pm 0.01$ fm will at least 
answer the question of whether the slope $y$ is steep or gentle.  

\begin{figure}
\begin{center}
\epsfig{file=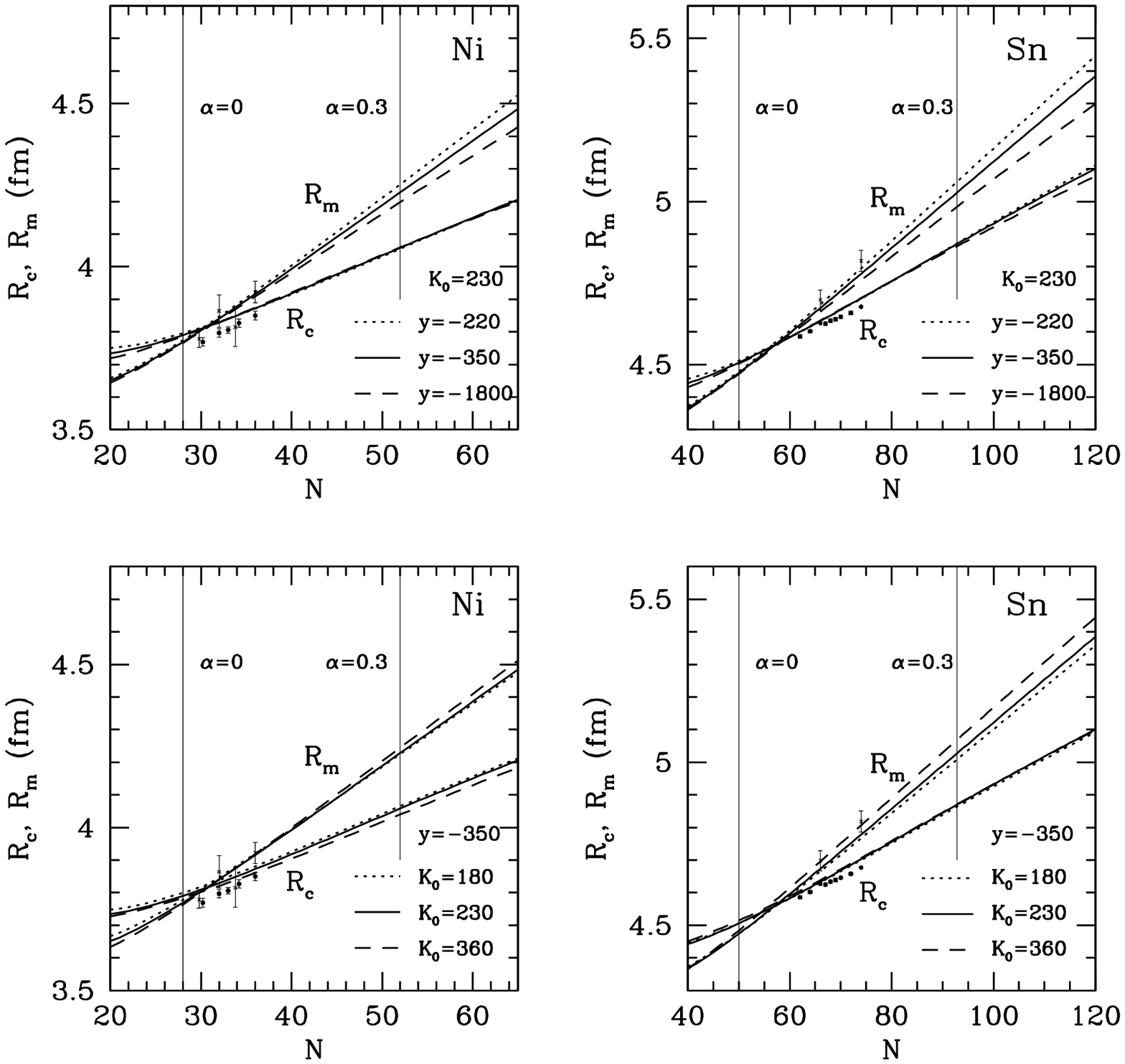,height=13cm}
\end{center}

\begin{flushleft}
{\small
Fig.\ 4.  The root-mean-square charge and matter radii of Ni and Sn isotopes
for combinations of $y=-220$, -350, -1800 MeV fm$^3$ and $K_0=180$, 230, 360 
MeV.  Experimental data on the root-mean-square charge radii (dots) and matter
radii (crosses) are taken from Refs.\ \cite{VJV} and \cite{Batty}, 
respectively.}
\end{flushleft}

\end{figure}

\begin{figure}
\begin{center}
\epsfig{file=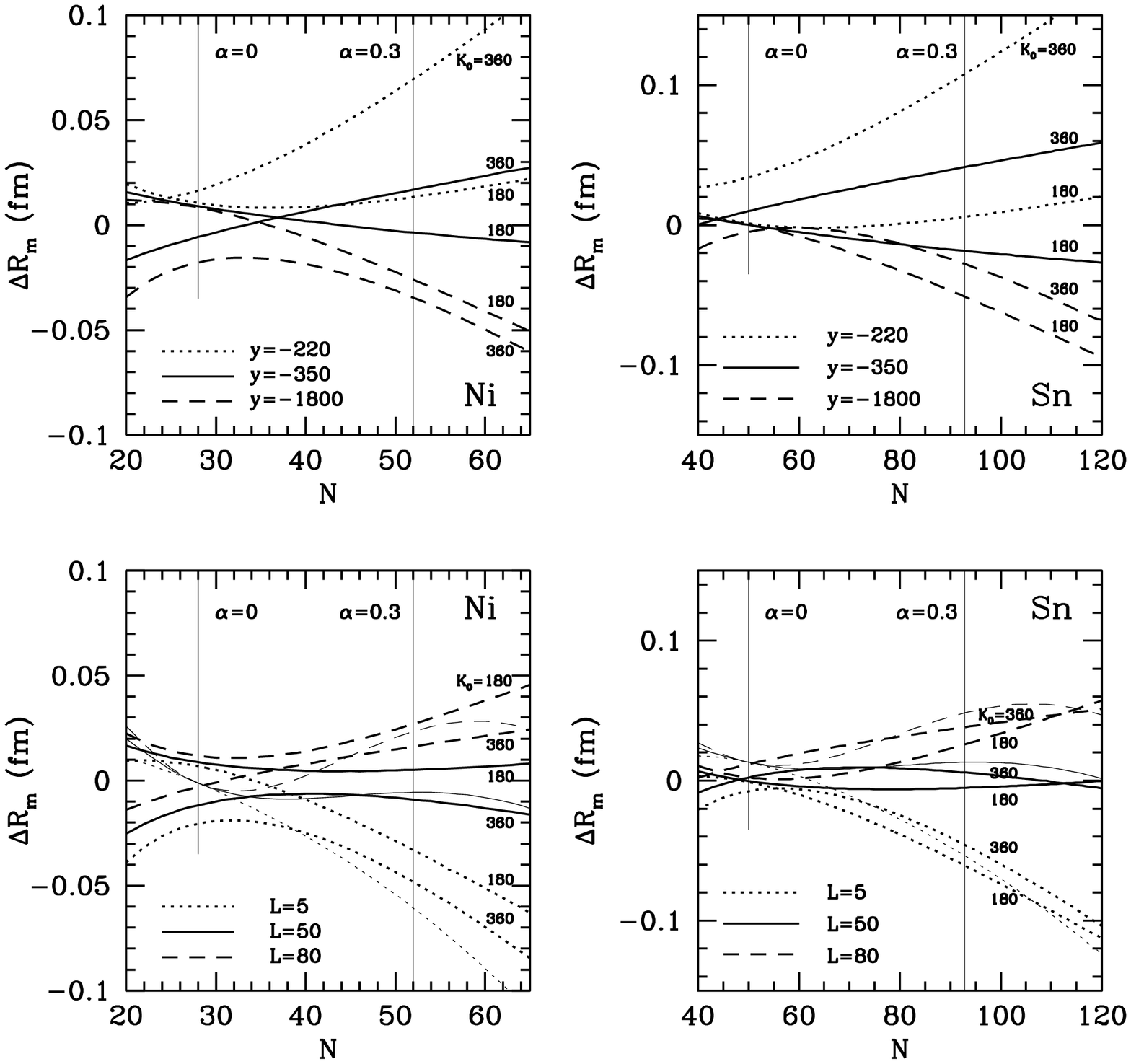,height=13cm}
\end{center}

\begin{flushleft}
{\small
Fig.\ 5.  Differences of the root-mean-square matter radii of Ni and Sn 
isotopes, $\Delta R_m$, from those calculated for $y=-350$ MeV fm$^3$ and 
$K_0=230$ MeV (upper panels) and for $L=50$ MeV and $K_0=230$ MeV (lower 
panels).  The thin curves in the lower panels are from formulas (\ref{rpfit}) 
and (\ref{rnfit}).}
\end{flushleft}

\end{figure}

     Several remarks, however, are needed here.  First, it is to be noted that
the experimental matter radii that can be derived from measurements of 
interaction cross sections \cite{Ozawa} and elastic scattering of protons and 
alpha particles \cite{Batty} depend strongly on treatment of the optical 
potential, in contrast to the charge radii that can be determined from elastic
electron scattering \cite{VJV,FB}, muonic X-ray experiments \cite{VJV,FB} and 
isotope-shift measurements \cite{Huber,FB}.  This dependence contributes to 
intrinsic uncertainties in the derived matter radii, which are typically of 
order or greater than $\pm 0.05$ fm \cite{Batty,Ray}.  Second, we recall that 
the present calculations of the radii $R_m$ and $R_c$ ignore the tails of the 
nucleon distributions arising from quantum-mechanical effects; the absence of 
such tails tends to reduce the radii.  Third, there are uncertainties in the 
calculated radii due to the absence of shell and pairing effects in
the present macroscopic models.  These models, which are fitted to 
the charge radii of nuclei on the {\em smoothed} $\beta$ stability line, 
provide the proton-closed-shell nuclei, Ni and Sn, with larger charge radii 
than the empirical values, as shown in Fig.\ 4.  Fourth, we can see from the
lower panels of Fig.\ 4 that with $K_0$ increased and $y$ fixed, the matter 
radii $R_m$ increase, a feature that prevents a clear derivation of $y$ from 
$R_m$.

     Such $K_0$ dependence of the matter radii $R_m$ is due mainly to the $K_0$
dependence of the saturation density, $n_s$, given by Eq.\ (\ref{ns}).  For 
fixed $y$, the $K_0$ dependence of $n_s$ is dominated by $n_0$, which 
increases with decreasing $K_0$ as in Fig.\ 1d.  Generally, this increase in 
$n_0$ tends to increase the inner nucleon densities (see the lower panels of 
Fig.\ 3) and hence reduce the radii $R_m$.  Note, however, that there is an 
opposite effect of $K_0$ on $R_m$.  This effect comes from the fact that a 
significant part of the nucleons are present in the deepest region of the 
nuclear surface where $r^2 n_n(r)$ and $r^2 n_p(r)$ are peaked.  In this 
region, the nucleon densities begin to drop in such a way that with decreasing
$K_0$, the surface diffuses further away.  This diffuseness, which can also be
seen from the lower panels of Fig.\ 3, is consistent with the results from 
microscopic nuclear models \cite{YST}.

      A cancellation between those counteracting effects is favoured for the 
purpose of deriving information about the saturation properties in a way 
independent of $K_0$.  It turns out that better cancellation can be achieved 
if we calculate, as in the lower panels of Fig.\ 6, the matter radii of Ni and
Sn isotopes for the EOS parameters optimized under fixed $L$ rather than $y$.
The charge radii likewise calculated are also nearly independent of $K_0$.  On
the other hand, under fixed $K_0$, the dependence of the matter and charge 
radii on $L$, as depicted in the upper panels of Fig.\ 6, is the same as that 
on $y$.  These results from Fig.\ 6 may open an opportunity to empirically 
determine the density symmetry coefficient $L$ from the isotopic dependence of
matter radii.

\begin{figure}
\begin{center}
\epsfig{file=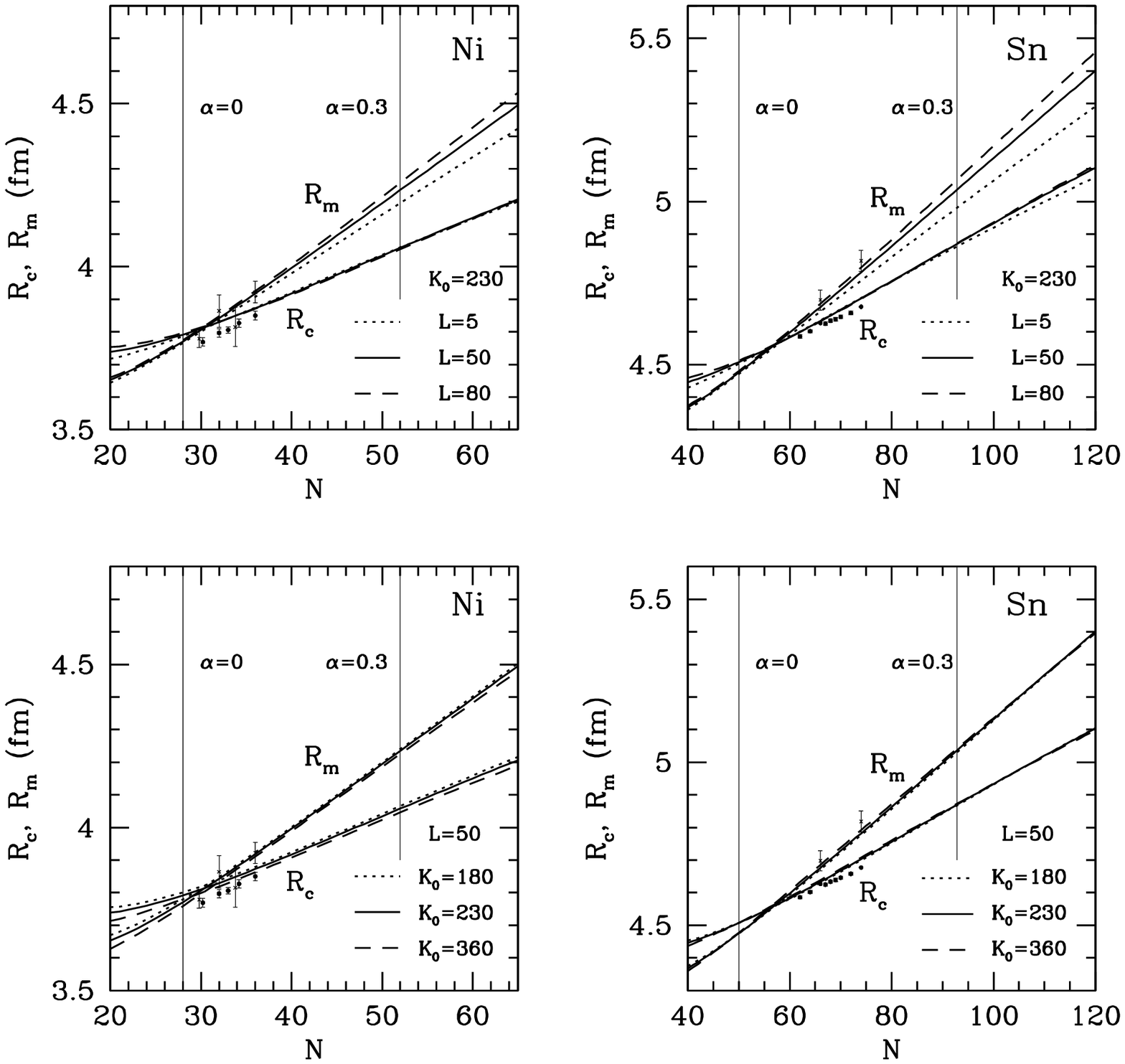,height=13cm}
\end{center}

\begin{flushleft}
{\small
Fig.\ 6.  Same as Fig.\ 4 for combinations of $L=5$, 50, 80 MeV and $K_0=180$,
230, 360 MeV.}
\end{flushleft}

\end{figure}

\begin{figure}
\begin{center}
\epsfig{file=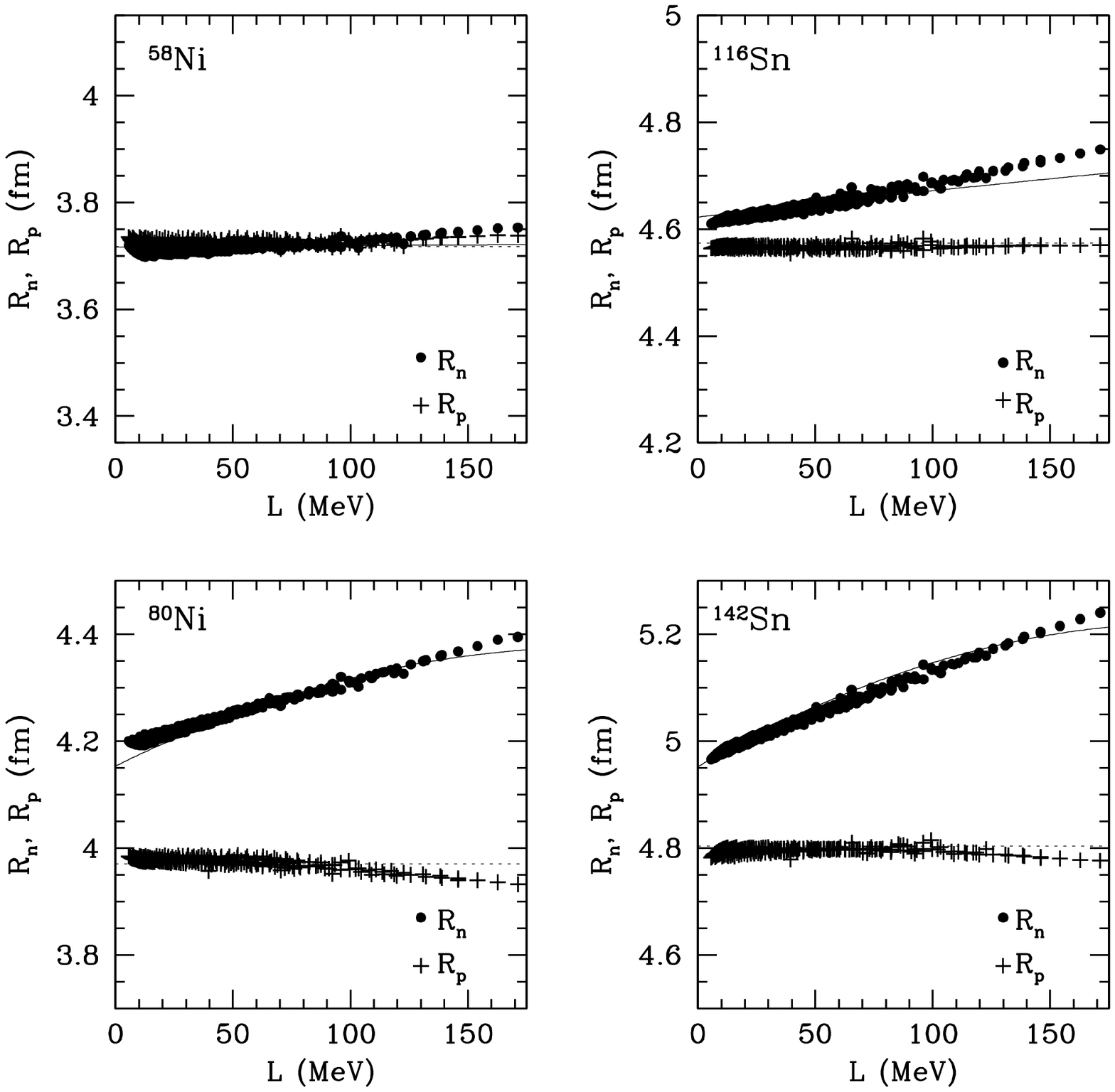,height=13cm}
\end{center}

\begin{flushleft}
{\small
Fig.\ 7.  The root-mean-square neutron (dots) and proton (crosses) radii of 
$^{58}$Ni, $^{80}$Ni, $^{116}$Sn and $^{142}$Sn for the EOS parameter sets 
tabulated in Table A.1.  The solid and dotted lines are from formulas 
(\ref{rnfit}) and (\ref{rpfit}), respectively.}
\end{flushleft}

\end{figure}

     For the purpose of parametrizing the matter and charge radii as functions
of $A$, $\alpha$ and $L$, it is useful to first obtain fitting formulas for 
the root-mean-square neutron and proton radii, defined as
\begin{equation}
 R_n=\left[N^{-1}\int d^3 r r^2 n_n(r)\right]^{1/2},~~~ 
 R_p=\left[Z^{-1}\int d^3 r r^2 n_p(r)\right]^{1/2}.
 \label{rnrp}
\end{equation}
To be fitted to are the radii $R_n$ and $R_p$ of Ni, Sn and Pb isotopes ranging
$0\leq\alpha\leq0.3$ $(0.35\leq x\leq0.5)$ as calculated from the present 
macroscopic model for the EOS parameter sets tabulated in Table A.1.  In Fig.\
7 the results for $^{56}$Ni, $^{80}$Ni, $^{116}$Sn and $^{142}$Ni are plotted 
as functions of $L$.  We observe from this figure that no uncertainties in 
$R_p$ more than $\pm0.03$ fm arise from the various values of $K_0$ and $L$, 
while the $K_0$ dependence of $R_n$ can be ignored as compared with the $L$ 
dependence, which is stronger at larger neutron excess.  These features, 
consistent with the dependence of $R_m$ and $R_c$ on $L$ and $K_0$ as 
mentioned above, lead us to the formulas,
\begin{equation}
R_p=c_1 A^{1/3}+c_2+c_3(\alpha-\alpha_0)^2,
\label{rpfit}
\end{equation}
with $c_1=0.914961$ fm, $c_2=-0.102372$ fm, $c_3=0.388905$ fm and $\alpha_0=
0.879704$, and
\begin{equation}
R_n=c_4 A^{1/3}(1+c_5 L\alpha^2+c_6 L^2\alpha^4)+c_7+c_8\alpha,
\label{rnfit}
\end{equation}
with $c_4=0.880489$ fm, $c_5=0.00635080$ MeV$^{-1}$, $c_6=-0.000172275$ 
MeV$^{-2}$, $c_7=0.301616$ fm and $c_8=0.193326$ fm.  Expressions 
(\ref{rpfit}) and (\ref{rnfit}) reproduce the original values with the 
root-mean-square deviations of 0.010 fm and 0.013 fm, respectively, and 
roughly obey a usual $A^{1/3}$ law for fixed $L$ and $\alpha$.  Finally, the 
parametrization of the matter and charge radii can be constructed from Eqs.\ 
(\ref{rpfit}) and (\ref{rnfit}) via the relations $R_m^2=(Z/A)R_p^2+(N/A)R_n^2
+3a_p^2/2$ and $R_c^2=R_p^2+3a_p^2/2$ that can be derived from Eqs.\ 
(\ref{Rc}), (\ref{Rm}) and (\ref{rnrp}).

      The lower panels of Fig.\ 5 depict the matter radii of Ni and Sn 
isotopes calculated for various combinations of $L$ and $K_0$ relative to 
those calculated for $L=50$ MeV and $K_0=230$ MeV.  As can be seen from these 
panels, $K_0$-induced uncertainties in the matter radii at constant $L$ are 
limited to the order of $\pm 0.015$ fm in the neutron-rich side, small enough 
for the $L$ dependence of the matter radii to be seen clearly.  This may play 
a role in determining $L$ from measurements of the matter radii.  Note, 
however, that possible data on matter radii for isotopes of a specific element
would be insufficient to determine the density symmetry coefficient $L$, as 
inferred from the differences between the empirical and calculated charge 
radii shown for Ni and Sn in Fig.\ 6.  In view of such differences as well as 
intrinsic uncertainties in the matter radii that can be deduced from proton and
alpha-particle elastic scattering data, systematics of the isotopic dependence
of matter radii would be required.

     It is interesting to note that the difference between the neutron and 
proton radii increases with increasing $L$ in a way dependent on neutron 
excess but almost independent of $K_0$ (see Fig.\ 7).  This leads us to study 
the isotopic dependence of neutron skin thickness in order to derive the 
density symmetry coefficient $L$.  The previous investigation \cite{OTSST} has
already suggested the possibility that the EOS of asymmetric nuclear matter 
will be probed by future detection of neutron skin thickness in unstable 
neutron-enriched nuclei.  However, the neutron skin structure depends strongly
on the EOS of nuclear matter at large neutron excess and low density \cite{B},
which is empirically hard to determine.  Moreover, the fact \cite{PR} that the
neutron skin thickness is thermodynamically relevant to the dependence of the 
surface tension on neutron excess complicates an extraction of the bulk 
properties of asymmetric nuclear matter; the connection between the surface 
and bulk properties remains to be clarified.  We will elsewhere reinvestigate 
how the neutron skin thickness is phenomenologically related to the EOS of 
nearly symmetric nuclear matter \cite{II}.

\section{Conclusion}
\label{sec:conc}

     In this paper we have derived the relations between the parameters 
characterizing the EOS of nearly symmetric nuclear matter from experimental 
data on radii and masses of stable nuclei.  We have found the linear relation 
(\ref{s0p0}) between the parameters $L$ and $S_0$ associated with the symmetry
energy.  It is interesting to regard the band structure on the $K_0$ versus 
$L$ plane (see Fig.\ 1c) as an empirically allowed region of $K_0$ and $L$.  
Future systematic measurements of the root-mean-square matter radii of 
unstable neutron-rich nuclei may narrow the allowed region in such a way as to
fix $L$ almost independently of $K_0$.  Once $L$ is determined, the slope $y$ 
would be given as a function of $K_0$.  It is instructive to note that the 
incompressibility was estimated to be $K_0=231 \pm5$ MeV from the observations
of the giant monopole resonances \cite{YCL}.  This estimate can stringently 
limit the allowed region, although it is model-dependent in the sense that it 
involves microscopic calculations of the resonance energies from a specific 
model of the effective nucleon-nucleon interactions \cite{BBDG}.

\section*{Acknowledgements}

    We are grateful to I. Tanihata, T. Suda, A. Ozawa, A. Kohama, H. Nakada, 
Y. Mochizuki and H. Koura for helpful discussions.  Author KO would like to 
thank Y. Mochizuki for hospitality during the course of this research at 
RIKEN.  This work was supported in part by RIKEN Special Postdoctoral 
Researchers Grant No.\ 011-52040.

\section*{Appendix: Evaluation of the parameters in Eq.\ (\ref{eos0})}

   In this appendix we tabulate (Table A.1) the optimal values of various EOS
parameters in Eq.\ (\ref{eos0}).

\newpage

\begin{table}[ht]
\begin{center}
\begin{minipage}[t]{10cm}
\begin{flushleft}
Table 1
\end{flushleft}
\noindent
The saturation properties of nuclear matter obtained from various
effective forces
\end{minipage}
\end{center}
\begin{center}
\begin{tabular}{ccccccc} \hline
 Force &  $n_0$  & $-w_0$  & $K_0$  & $S_0$  & $L$ & $-y$ \\
\hline
  SI   &   0.155  & 16.0  &  371 & 29.2 &  1.18 & 19700\\
  SII  &   0.148  & 16.0  &  342 & 34.2 &  50.0 &   524\\
 SIII  &   0.145  & 15.9  &  356 & 28.2 &  9.87 &  2330\\
 SIV   &   0.151  & 16.0  &  325 & 31.2 &  63.5 &   353\\
 SV    &   0.155  & 16.1  &  306 & 32.8 &  96.1 &   224\\
 SVI   &   0.144  & 15.8  &  364 & 26.9 & $-7.38$ & $-3080$\\
 Ska   &   0.155  & 16.0  &  263 & 32.9 &  74.6 &   249\\
 Skb   &   0.155  & 16.0  &  263 & 23.9 &  47.5 &   284\\
 SG-0  &   0.168  & 16.7  &  253 & 35.6 &  41.6 &   430\\
 SGI   &   0.154  & 15.8  &  261 & 28.3 &  64.1 &   250\\
 SGII  &   0.158  & 15.6  &  215 & 26.8 &  37.6 &   322\\
 SkM   &   0.160  & 15.8  &  217 & 30.7 &  49.3 &   281\\
 SkM$^*$ & 0.160  & 15.8  &  217 & 30.0 &  45.8 &   296\\
 $E$   &   0.159  & 16.1  &  334 & 27.6 & $-31.3$ &  $-617$\\
 $E_\sigma$ &   0.163  & 16.0  &  249 & 26.4 & $-36.9$ & $-364$\\
 $Z$   &   0.159  & 16.0  &  330  & 26.8  & $-49.8$ & $-373$\\
 $Z_\sigma$ &   0.163  & 15.9  &  233 & 26.7 & $-29.4$ & $-432$\\
 $Z_\sigma^*$ & 0.163  & 16.0  &  235 & 28.8 & $-4.58$ & $-3030$\\
 $R_\sigma$ &   0.158  & 15.6  &  238 & 30.6 &  85.7 & 179\\
 $G_\sigma$ &   0.158  & 15.6  &  237 & 31.4 &  94.0 & 167\\
 MSkA  &   0.154  & 16.0  &  314 & 30.4 &  57.2 & 361\\
 SkT6  &   0.161  & 16.0  &  236 & 30.0 &  30.8 & 475\\
 SkP   &   0.163  & 16.0  &  201 & 30.0 &  19.5 & 632\\
 SkSC4 &   0.161  & 15.9  &  235 & 28.8 & $-2.17$ & $-6460$\\
 SkX   &   0.155  & 16.1  &  271 & 31.1 &  33.2 & 545\\
 MSk7  &   0.158  & 15.8  &  231 & 27.9 &  9.36 & 1460\\
 BSk1  &   0.157  & 15.8  &  231 & 27.8 &  7.15 & 1908\\
 SLy4  &   0.160  & 16.0  &  230 & 32.0 &  45.9 & 335\\
 SLy7  &   0.158  & 15.9  &  230 & 32.0 &  47.2 & 328\\
\hline
 TM1   &   0.145    & 16.3    &  281    & 37.9   &  114 & 215\\
\hline
\end{tabular}
\end{center} 
\end{table}  

\newpage

\begin{table}[ht]
\begin{center}
\begin{minipage}[t]{10cm}
\begin{flushleft}
Table A.1
\end{flushleft}
\noindent
Optimal values of the parameters $n_0$, $w_0$, $S_0$, $L$ and $F_0$ for 
fixed $y$ and $K_0$
\end{minipage}
\end{center}
\begin{center}
\begin{tabular}{ccccccc} \hline
  $-y$  & $K_0$  & $n_0$  & $-w_0$  & $S_0$ & $L$ & $F_0$ \\
\hline
         200 &         180 &  0.16931     &   16.257     &   32.899     &
   58.296     &   71.300     \\
         200 &         190 &  0.16695     &   16.239     &   33.154     &
   62.887     &   70.892     \\
         200 &         200 &  0.16491     &   16.227     &   33.430     &
   67.572     &   70.534     \\
         200 &         210 &  0.16298     &   16.214     &   33.723     &
   72.421     &   70.221     \\
         200 &         220 &  0.16125     &   16.204     &   34.037     &
   77.398     &   69.914     \\
         200 &         230 &  0.15965     &   16.193     &   34.365     &
   82.511     &   69.539     \\
         200 &         240 &  0.15793     &   16.120     &   34.543     &
   87.488     &   67.031     \\
         200 &         250 &  0.15685     &   16.186     &   35.113     &
   93.278     &   69.320     \\
         200 &         260 &  0.15228     &   15.932     &   34.905     &
   99.328     &   63.190     \\
         200 &         270 &  0.15200     &   15.870     &   34.933     &
   103.42     &   60.072     \\
         200 &         280 &  0.15298     &   16.157     &   36.349     &
   110.88     &   68.403     \\
         200 &         290 &  0.15184     &   16.153     &   36.856     &
   117.32     &   68.297     \\
         200 &         300 &  0.14880     &   15.858     &   36.503     &
   122.66     &   59.622     \\
         200 &         310 &  0.14972     &   16.144     &   37.939     &
   130.93     &   68.070     \\
         200 &         320 &  0.14867     &   16.139     &   38.540     &
   138.26     &   68.006     \\
         200 &         330 &  0.14772     &   16.135     &   39.190     &
   145.92     &   67.944     \\
         200 &         340 &  0.14673     &   16.131     &   39.904     &
   154.10     &   67.911     \\
         200 &         350 &  0.14582     &   16.130     &   40.684     &
   162.75     &   67.969     \\
         200 &         360 &  0.14444     &   16.056     &   41.232     &
   171.28     &   65.851     \\
         220 &         180 &  0.16925     &   16.250     &   32.350     &
   52.129     &   71.133     \\
         220 &         190 &  0.16698     &   16.234     &   32.564     &
   56.141     &   70.709     \\
         220 &         200 &  0.16489     &   16.218     &   32.791     &
   60.264     &   70.320     \\
         220 &         210 &  0.16303     &   16.207     &   33.031     &
   64.467     &   69.959     \\
         220 &         220 &  0.16131     &   16.196     &   33.286     &
   68.784     &   69.625     \\
         220 &         230 &  0.15969     &   16.184     &   33.550     &
   73.214     &   69.276     \\
         220 &         240 &  0.15821     &   16.175     &   33.835     &
   77.767     &   68.969     \\
         220 &         250 &  0.15748     &   16.184     &   34.183     &
   82.224     &   68.520     \\
         220 &         260 &  0.15412     &   16.053     &   34.231     &
   87.496     &   65.686     \\
         220 &         270 &  0.15438     &   16.179     &   34.850     &
   92.349     &   69.206     \\
\hline
\end{tabular}
\end{center} 
\end{table}  

\newpage

\begin{table}[ht]
\begin{center}
\begin{minipage}[t]{10cm}
\begin{flushleft}
Table A.1---continued
\end{flushleft}
\end{minipage}
\end{center}
\begin{center}
\begin{tabular}{ccccccc} \hline
  $-y$  & $K_0$  & $n_0$  & $-w_0$  & $S_0$ & $L$ & $F_0$ \\
\hline
         220 &         280 &  0.15265     &   16.003     &   34.543     &
   96.002     &   63.371     \\
         220 &         290 &  0.15128     &   16.019     &   35.173     &
   102.16     &   63.971     \\
         220 &         300 &  0.15110     &   16.133     &   35.917     &
   108.05     &   67.413     \\
         220 &         310 &  0.15019     &   16.135     &   36.354     &
   113.69     &   67.388     \\
         220 &         320 &  0.14856     &   16.000     &   36.340     &
   118.60     &   63.115     \\
         220 &         330 &  0.14832     &   16.126     &   37.295     &
   125.72     &   67.102     \\
         220 &         340 &  0.14742     &   16.123     &   37.813     &
   132.14     &   67.039     \\
         220 &         350 &  0.14651     &   16.128     &   38.395     &
   138.97     &   67.357     \\
         220 &         360 &  0.14570     &   16.114     &   38.963     &
   145.86     &   66.841     \\
         250 &         180 &  0.16876     &   16.191     &   31.643     &
   45.001     &   69.337     \\
         250 &         190 &  0.16694     &   16.225     &   31.893     &
   48.399     &   70.488     \\
         250 &         200 &  0.16490     &   16.210     &   32.071     &
   51.862     &   70.075     \\
         250 &         210 &  0.16305     &   16.197     &   32.251     &
   55.383     &   69.669     \\
         250 &         220 &  0.16136     &   16.184     &   32.449     &
   58.988     &   69.240     \\
         250 &         230 &  0.15978     &   16.174     &   32.655     &
   62.675     &   68.890     \\
         250 &         240 &  0.15829     &   16.154     &   32.851     &
   66.410     &   68.193     \\
         250 &         250 &  0.15707     &   16.155     &   33.098     &
   70.239     &   68.083     \\
         250 &         260 &  0.15590     &   16.159     &   33.388     &
   74.246     &   68.228     \\
         250 &         270 &  0.15479     &   16.129     &   33.574     &
   78.083     &   66.874     \\
         250 &         280 &  0.15364     &   16.142     &   33.882     &
   82.331     &   67.477     \\
         250 &         290 &  0.15087     &   15.950     &   33.612     &
   86.147     &   62.109     \\
         250 &         300 &  0.15173     &   16.132     &   34.449     &
   90.818     &   66.855     \\
         250 &         310 &  0.14900     &   16.081     &   34.601     &
   95.985     &   67.159     \\
         250 &         320 &  0.14894     &   16.028     &   34.781     &
   99.635     &   63.947     \\
         250 &         330 &  0.14889     &   16.113     &   35.403     &
   104.62     &   66.202     \\
         250 &         340 &  0.14818     &   16.074     &   35.654     &
   109.08     &   64.564     \\
         250 &         350 &  0.14727     &   16.105     &   36.144     &
   114.54     &   65.893     \\
         250 &         360 &  0.14653     &   16.103     &   36.548     &
   119.72     &   65.769     \\
\hline
\end{tabular}
\end{center} 
\end{table}  

\newpage

\begin{table}[ht]
\begin{center}
\begin{minipage}[t]{10cm}
\begin{flushleft}
Table A.1---continued
\end{flushleft}
\end{minipage}
\end{center}
\begin{center}
\begin{tabular}{ccccccc} \hline
  $-y$  & $K_0$  & $n_0$  & $-w_0$  & $S_0$ & $L$ & $F_0$ \\
\hline
         300 &         180 &  0.16915     &   16.232     &   31.027     &
   36.685     &   70.687     \\
         300 &         190 &  0.16679     &   16.206     &   31.127     &
   39.398     &   69.997     \\
         300 &         200 &  0.16492     &   16.201     &   31.261     &
   42.123     &   69.789     \\
         300 &         210 &  0.16304     &   16.185     &   31.387     &
   44.919     &   69.318     \\
         300 &         220 &  0.16138     &   16.172     &   31.522     &
   47.747     &   68.875     \\
         300 &         230 &  0.15991     &   16.164     &   31.669     &
   50.612     &   68.492     \\
         300 &         240 &  0.15845     &   16.152     &   31.816     &
   53.547     &   68.116     \\
         300 &         250 &  0.15716     &   16.144     &   31.977     &
   56.518     &   67.747     \\
         300 &         260 &  0.15594     &   16.135     &   32.136     &
   59.533     &   67.382     \\
         300 &         270 &  0.15487     &   16.129     &   32.312     &
   62.595     &   67.030     \\
         300 &         280 &  0.15386     &   16.128     &   32.498     &
   65.712     &   66.862     \\
         300 &         290 &  0.15301     &   16.134     &   32.722     &
   68.907     &   66.850     \\
         300 &         300 &  0.15309     &   16.206     &   33.118     &
   72.109     &   68.333     \\
         300 &         310 &  0.15101     &   16.103     &   33.093     &
   75.485     &   65.682     \\
         300 &         320 &  0.14987     &   16.058     &   33.149     &
   78.644     &   64.466     \\
         300 &         330 &  0.14883     &   16.071     &   33.480     &
   82.484     &   65.001     \\
         300 &         340 &  0.14699     &   15.926     &   33.240     &
   85.428     &   60.873     \\
         300 &         350 &  0.14680     &   15.964     &   33.630     &
   89.092     &   61.549     \\
         300 &         360 &  0.14746     &   16.025     &   34.068     &
   92.411     &   62.205     \\
         350 &         180 &  0.16905     &   16.224     &   30.543     &
   30.974     &   70.513     \\
         350 &         190 &  0.16677     &   16.201     &   30.617     &
   33.221     &   69.869     \\
         350 &         200 &  0.16472     &   16.180     &   30.696     &
   35.495     &   69.261     \\
         350 &         210 &  0.16297     &   16.174     &   30.811     &
   37.813     &   69.037     \\
         350 &         220 &  0.16086     &   16.131     &   30.837     &
   40.166     &   67.933     \\
         350 &         230 &  0.15979     &   16.145     &   31.002     &
   42.498     &   67.975     \\
         350 &         240 &  0.15877     &   16.167     &   31.189     &
   44.902     &   68.422     \\
         350 &         250 &  0.15729     &   16.138     &   31.258     &
   47.315     &   67.474     \\
         350 &         260 &  0.15608     &   16.128     &   31.374     &
   49.775     &   67.057     \\
         350 &         270 &  0.15500     &   16.121     &   31.503     &
   52.263     &   66.690     \\
\hline
\end{tabular}
\end{center} 
\end{table}  

\newpage

\begin{table}[ht]
\begin{center}
\begin{minipage}[t]{10cm}
\begin{flushleft}
Table A.1---continued
\end{flushleft}
\end{minipage}
\end{center}
\begin{center}
\begin{tabular}{ccccccc} \hline
  $-y$  & $K_0$  & $n_0$  & $-w_0$  & $S_0$ & $L$ & $F_0$ \\
\hline
         350 &         280 &  0.15412     &   16.143     &   31.701     &
   54.852     &   67.309     \\
         350 &         290 &  0.15305     &   16.108     &   31.777     &
   57.343     &   65.982     \\
         350 &         300 &  0.15211     &   16.101     &   31.917     &
   59.953     &   65.635     \\
         350 &         310 &  0.15133     &   16.099     &   32.071     &
   62.569     &   65.373     \\
         350 &         320 &  0.15117     &   16.227     &   32.539     &
   65.599     &   69.348     \\
         350 &         330 &  0.14978     &   16.088     &   32.384     &
   67.950     &   64.685     \\
         350 &         340 &  0.14959     &   16.073     &   32.489     &
   70.326     &   63.401     \\
         350 &         350 &  0.14785     &   16.073     &   32.714     &
   73.757     &   64.408     \\
         350 &         360 &  0.14741     &   16.025     &   32.734     &
   76.137     &   62.538     \\
         400 &         180 &  0.16902     &   16.219     &   30.206     &
   26.807     &   70.423     \\
         400 &         190 &  0.16675     &   16.200     &   30.263     &
   28.736     &   69.895     \\
         400 &         200 &  0.16446     &   16.164     &   30.287     &
   30.693     &   69.005     \\
         400 &         210 &  0.16303     &   16.173     &   30.412     &
   32.645     &   68.967     \\
         400 &         220 &  0.16138     &   16.158     &   30.481     &
   34.628     &   68.483     \\
         400 &         230 &  0.15988     &   16.148     &   30.566     &
   36.642     &   68.057     \\
         400 &         240 &  0.15855     &   16.138     &   30.654     &
   38.668     &   67.603     \\
         400 &         250 &  0.15728     &   16.130     &   30.746     &
   40.727     &   67.250     \\
         400 &         260 &  0.15615     &   16.122     &   30.845     &
   42.801     &   66.850     \\
         400 &         270 &  0.15505     &   16.113     &   30.942     &
   44.900     &   66.429     \\
         400 &         280 &  0.15406     &   16.107     &   31.043     &
   47.016     &   66.044     \\
         400 &         290 &  0.15315     &   16.100     &   31.156     &
   49.162     &   65.654     \\
         400 &         300 &  0.15227     &   16.095     &   31.265     &
   51.332     &   65.344     \\
         400 &         310 &  0.15146     &   16.090     &   31.381     &
   53.526     &   65.004     \\
         400 &         320 &  0.15070     &   16.087     &   31.508     &
   55.754     &   64.721     \\
         400 &         330 &  0.14996     &   16.081     &   31.627     &
   58.000     &   64.364     \\
         400 &         340 &  0.14938     &   16.164     &   31.976     &
   60.648     &   67.165     \\
         400 &         350 &  0.14854     &   16.071     &   31.895     &
   62.628     &   63.808     \\
         400 &         360 &  0.14701     &   16.095     &   32.082     &
   65.471     &   65.588     \\
\hline
\end{tabular}
\end{center} 
\end{table}  

\newpage

\begin{table}[ht]
\begin{center}
\begin{minipage}[t]{10cm}
\begin{flushleft}
Table A.1---continued
\end{flushleft}
\end{minipage}
\end{center}
\begin{center}
\begin{tabular}{ccccccc} \hline
  $-y$  & $K_0$  & $n_0$  & $-w_0$  & $S_0$ & $L$ & $F_0$ \\
\hline
         500 &         180 &  0.16895     &   16.213     &   29.750     &
   21.131     &   70.289     \\
         500 &         190 &  0.16674     &   16.194     &   29.786     &
   22.628     &   69.745     \\
         500 &         200 &  0.16506     &   16.201     &   29.868     &
   24.127     &   69.843     \\
         500 &         210 &  0.16324     &   16.179     &   29.901     &
   25.644     &   69.044     \\
         500 &         220 &  0.16136     &   16.149     &   29.914     &
   27.191     &   68.228     \\
         500 &         230 &  0.15991     &   16.141     &   29.972     &
   28.739     &   67.851     \\
         500 &         240 &  0.15856     &   16.130     &   30.028     &
   30.302     &   67.377     \\
         500 &         250 &  0.15731     &   16.120     &   30.085     &
   31.875     &   66.940     \\
         500 &         260 &  0.15620     &   16.112     &   30.152     &
   33.459     &   66.504     \\
         500 &         270 &  0.15513     &   16.104     &   30.216     &
   35.060     &   66.079     \\
         500 &         280 &  0.15421     &   16.098     &   30.288     &
   36.663     &   65.685     \\
         500 &         290 &  0.15330     &   16.092     &   30.359     &
   38.286     &   65.293     \\
         500 &         300 &  0.15243     &   16.086     &   30.437     &
   39.934     &   64.956     \\
         500 &         310 &  0.15165     &   16.080     &   30.511     &
   41.581     &   64.550     \\
         500 &         320 &  0.15091     &   16.076     &   30.594     &
   43.249     &   64.240     \\
         500 &         330 &  0.15022     &   16.073     &   30.678     &
   44.930     &   63.950     \\
         500 &         340 &  0.14957     &   16.069     &   30.765     &
   46.624     &   63.597     \\
         500 &         350 &  0.14892     &   16.063     &   30.852     &
   48.341     &   63.250     \\
         500 &         360 &  0.14912     &   16.222     &   31.314     &
   50.399     &   68.034     \\
         600 &         180 &  0.16890     &   16.208     &   29.461     &
   17.443     &   70.159     \\
         600 &         190 &  0.16669     &   16.188     &   29.475     &
   18.664     &   69.602     \\
         600 &         200 &  0.16470     &   16.172     &   29.496     &
   19.899     &   69.132     \\
         600 &         210 &  0.16393     &   16.245     &   29.692     &
   21.132     &   70.991     \\
         600 &         220 &  0.16137     &   16.148     &   29.560     &
   22.389     &   68.204     \\
         600 &         230 &  0.15990     &   16.135     &   29.596     &
   23.651     &   67.684     \\
         600 &         240 &  0.15856     &   16.123     &   29.634     &
   24.920     &   67.189     \\
         600 &         250 &  0.15735     &   16.114     &   29.675     &
   26.194     &   66.714     \\
         600 &         260 &  0.15626     &   16.108     &   29.724     &
   27.476     &   66.344     \\
         600 &         270 &  0.15519     &   16.097     &   29.766     &
   28.770     &   65.856     \\
\hline
\end{tabular}
\end{center} 
\end{table}  

\newpage

\begin{table}[ht]
\begin{center}
\begin{minipage}[t]{10cm}
\begin{flushleft}
Table A.1---continued
\end{flushleft}
\end{minipage}
\end{center}
\begin{center}
\begin{tabular}{ccccccc} \hline
  $-y$  & $K_0$  & $n_0$  & $-w_0$  & $S_0$ & $L$ & $F_0$ \\
\hline
         600 &         280 &  0.15422     &   16.091     &   29.818     &
   30.077     &   65.479     \\
         600 &         290 &  0.15336     &   16.085     &   29.871     &
   31.381     &   65.049     \\
         600 &         300 &  0.15254     &   16.081     &   29.930     &
   32.701     &   64.714     \\
         600 &         310 &  0.15176     &   16.075     &   29.988     &
   34.031     &   64.339     \\
         600 &         320 &  0.15106     &   16.070     &   30.047     &
   35.362     &   63.953     \\
         600 &         330 &  0.15042     &   16.072     &   30.118     &
   36.708     &   63.752     \\
         600 &         340 &  0.14968     &   16.061     &   30.168     &
   38.071     &   63.287     \\
         600 &         350 &  0.15232     &   16.340     &   30.965     &
   39.528     &   69.329     \\
         600 &         360 &  0.14849     &   16.053     &   30.297     &
   40.808     &   62.629     \\
         800 &         180 &  0.16884     &   16.203     &   29.105     &
   12.929     &   70.077     \\
         800 &         190 &  0.16688     &   16.200     &   29.137     &
   13.823     &   69.926     \\
         800 &         200 &  0.16491     &   16.183     &   29.142     &
   14.726     &   69.348     \\
         800 &         210 &  0.16455     &   16.283     &   29.351     &
   15.607     &   71.847     \\
         800 &         220 &  0.16132     &   16.135     &   29.140     &
   16.558     &   67.801     \\
         800 &         230 &  0.15982     &   16.127     &   29.147     &
   17.478     &   67.534     \\
         800 &         240 &  0.15858     &   16.118     &   29.174     &
   18.397     &   67.030     \\
         800 &         250 &  0.15734     &   16.107     &   29.193     &
   19.327     &   66.550     \\
         800 &         260 &  0.15624     &   16.099     &   29.217     &
   20.258     &   66.079     \\
         800 &         270 &  0.15527     &   16.092     &   29.249     &
   21.193     &   65.645     \\
         800 &         280 &  0.15432     &   16.085     &   29.276     &
   22.133     &   65.237     \\
         800 &         290 &  0.15341     &   16.076     &   29.305     &
   23.083     &   64.764     \\
         800 &         300 &  0.15265     &   16.073     &   29.342     &
   24.028     &   64.400     \\
         800 &         310 &  0.15186     &   16.067     &   29.373     &
   24.983     &   64.001     \\
         800 &         320 &  0.15117     &   16.063     &   29.414     &
   25.943     &   63.644     \\
         800 &         330 &  0.15047     &   16.057     &   29.448     &
   26.909     &   63.268     \\
         800 &         340 &  0.14983     &   16.053     &   29.489     &
   27.881     &   62.933     \\
         800 &         350 &  0.14925     &   16.049     &   29.529     &
   28.852     &   62.575     \\
         800 &         360 &  0.14866     &   16.041     &   29.572     &
   29.839     &   62.127     \\
\hline
\end{tabular}
\end{center} 
\end{table}  

\newpage

\begin{table}[ht]
\begin{center}
\begin{minipage}[t]{10cm}
\begin{flushleft}
Table A.1---continued
\end{flushleft}
\end{minipage}
\end{center}
\begin{center}
\begin{tabular}{ccccccc} \hline
  $-y$  & $K_0$  & $n_0$  & $-w_0$  & $S_0$ & $L$ & $F_0$ \\
\hline
        1000 &         180 &  0.16891     &   16.213     &   28.929     &
   10.276     &   70.446     \\
        1000 &         190 &  0.16656     &   16.179     &   28.890     &
   10.985     &   69.460     \\
        1000 &         200 &  0.16460     &   16.164     &   28.888     &
   11.700     &   68.969     \\
        1000 &         210 &  0.16411     &   16.266     &   29.106     &
   12.415     &   71.720     \\
        1000 &         220 &  0.16127     &   16.133     &   28.888     &
   13.136     &   67.828     \\
        1000 &         230 &  0.15989     &   16.124     &   28.890     &
   13.853     &   67.378     \\
        1000 &         240 &  0.15857     &   16.114     &   28.907     &
   14.584     &   66.911     \\
        1000 &         250 &  0.15738     &   16.103     &   28.920     &
   15.314     &   66.392     \\
        1000 &         260 &  0.15626     &   16.095     &   28.933     &
   16.047     &   65.958     \\
        1000 &         270 &  0.15525     &   16.086     &   28.950     &
   16.783     &   65.496     \\
        1000 &         280 &  0.15433     &   16.080     &   28.969     &
   17.520     &   65.071     \\
        1000 &         290 &  0.15345     &   16.074     &   28.988     &
   18.261     &   64.677     \\
        1000 &         300 &  0.15268     &   16.069     &   29.012     &
   19.001     &   64.238     \\
        1000 &         310 &  0.15191     &   16.063     &   29.037     &
   19.752     &   63.864     \\
        1000 &         320 &  0.15121     &   16.059     &   29.060     &
   20.499     &   63.503     \\
        1000 &         330 &  0.15053     &   16.052     &   29.085     &
   21.253     &   63.069     \\
        1000 &         340 &  0.14991     &   16.048     &   29.110     &
   22.008     &   62.729     \\
        1000 &         350 &  0.14935     &   16.045     &   29.141     &
   22.763     &   62.395     \\
        1000 &         360 &  0.14880     &   16.042     &   29.171     &
   23.525     &   62.056     \\
        1400 &         180 &  0.16872     &   16.195     &   28.677     &
   7.2841     &   69.908     \\
        1400 &         190 &  0.16657     &   16.178     &   28.657     &
   7.7828     &   69.419     \\
        1400 &         200 &  0.16460     &   16.160     &   28.641     &
   8.2861     &   68.842     \\
        1400 &         210 &  0.16291     &   16.146     &   28.633     &
   8.7877     &   68.295     \\
        1400 &         220 &  0.16130     &   16.132     &   28.624     &
   9.2955     &   67.785     \\
        1400 &         230 &  0.15979     &   16.115     &   28.608     &
   9.8043     &   67.148     \\
        1400 &         240 &  0.15867     &   16.117     &   28.633     &
   10.311     &   66.953     \\
        1400 &         250 &  0.15735     &   16.097     &   28.615     &
   10.825     &   66.220     \\
        1400 &         260 &  0.15632     &   16.093     &   28.625     &
   11.336     &   65.868     \\
        1400 &         270 &  0.15527     &   16.084     &   28.627     &
   11.852     &   65.403     \\
\hline
\end{tabular}
\end{center} 
\end{table}  

\newpage

\begin{table}[ht]
\begin{center}
\begin{minipage}[t]{10cm}
\begin{flushleft}
Table A.1---continued
\end{flushleft}
\end{minipage}
\end{center}
\begin{center}
\begin{tabular}{ccccccc} \hline
  $-y$  & $K_0$  & $n_0$  & $-w_0$  & $S_0$ & $L$ & $F_0$ \\
\hline
        1400 &         280 &  0.15433     &   16.075     &   28.634     &
   12.369     &   64.909     \\
        1400 &         290 &  0.15352     &   16.071     &   28.645     &
   12.883     &   64.546     \\
        1400 &         300 &  0.15269     &   16.063     &   28.655     &
   13.405     &   64.064     \\
        1400 &         310 &  0.15196     &   16.058     &   28.667     &
   13.924     &   63.663     \\
        1400 &         320 &  0.15126     &   16.053     &   28.681     &
   14.447     &   63.317     \\
        1400 &         330 &  0.15065     &   16.048     &   28.693     &
   14.965     &   62.884     \\
        1400 &         340 &  0.15003     &   16.044     &   28.708     &
   15.490     &   62.512     \\
        1400 &         350 &  0.14941     &   16.039     &   28.713     &
   16.015     &   62.182     \\
        1400 &         360 &  0.14889     &   16.035     &   28.738     &
   16.544     &   61.778     \\
        1800 &         180 &  0.16868     &   16.193     &   28.552     &
   5.6423     &   69.913     \\
        1800 &         190 &  0.16652     &   16.174     &   28.525     &
   6.0275     &   69.339     \\
        1800 &         200 &  0.16460     &   16.158     &   28.508     &
   6.4145     &   68.779     \\
        1800 &         210 &  0.16281     &   16.142     &   28.488     &
   6.8046     &   68.254     \\
        1800 &         220 &  0.16126     &   16.129     &   28.475     &
   7.1940     &   67.714     \\
        1800 &         230 &  0.15983     &   16.116     &   28.465     &
   7.5857     &   67.193     \\
        1800 &         240 &  0.15846     &   16.104     &   28.446     &
   7.9788     &   66.736     \\
        1800 &         250 &  0.15738     &   16.098     &   28.459     &
   8.3715     &   66.256     \\
        1800 &         260 &  0.15629     &   16.090     &   28.457     &
   8.7665     &   65.790     \\
        1800 &         270 &  0.15530     &   16.082     &   28.457     &
   9.1619     &   65.332     \\
        1800 &         280 &  0.15437     &   16.074     &   28.459     &
   9.5591     &   64.869     \\
        1800 &         290 &  0.15349     &   16.065     &   28.457     &
   9.9567     &   64.391     \\
        1800 &         300 &  0.15271     &   16.058     &   28.462     &
   10.354     &   63.912     \\
        1800 &         310 &  0.15199     &   16.056     &   28.471     &
   10.754     &   63.584     \\
        1800 &         320 &  0.15129     &   16.050     &   28.478     &
   11.155     &   63.167     \\
        1800 &         330 &  0.15068     &   16.046     &   28.485     &
   11.553     &   62.780     \\
        1800 &         340 &  0.15009     &   16.043     &   28.498     &
   11.955     &   62.431     \\
        1800 &         350 &  0.14948     &   16.037     &   28.507     &
   12.361     &   62.062    \\
        1800 &         360 &  0.14895     &   16.032     &   28.515     &
   12.763    &   61.648    \\
\hline
\end{tabular}
\end{center} 
\end{table}  

\end{document}